\documentstyle[12pt]{article} 
\catcode`\@=11 
\def\marginnote#1{} 
\newcount\hour 
\newcount\minute 
\newtoks\amorpm 
\hour=\time\divide\hour by60 
\minute=\time{\multiply\hour by60 \global\advance\minute by-\hour} 
\edef\standardtime{{\ifnum\hour<12 \global\amorpm={am}%
        \else\global\amorpm={pm}\advance\hour by-12 \fi 
        \ifnum\hour=0 \hour=12 \fi 
        \number\hour:\ifnum\minute<10 0\fi\number\minute\the\amorpm}} 
\edef\militarytime{\number\hour:\ifnum\minute<10 0\fi\number\minute} 
\def\draftlabel#1{{\@bsphack\if@filesw {\let\thepage\relax 
   \xdef\@gtempa{\write\@auxout{\string 
      \newlabel{#1}{{\@currentlabel}{\thepage}}}}}\@gtempa 
   \if@nobreak \ifvmode\nobreak\fi\fi\fi\@esphack} 
        \gdef\@eqnlabel{#1}} 
\def\@eqnlabel{} 
\def\@vacuum{} 
\def\draftmarginnote#1{\marginpar{\raggedright\scriptsize\tt#1}} 
\def\draft{\oddsidemargin -.5truein 
        \def\@oddfoot{\sl preliminary draft \hfil 
        \rm\thepage\hfil\sl\today\quad\militarytime} 
        \let\@evenfoot\@oddfoot \overfullrule 3pt 
        \let\label=\draftlabel 
        \let\marginnote=\draftmarginnote 
   \def\@eqnnum{(\theequation)\rlap{\kern\marginparsep\tt\@eqnlabel}%
\global\let\@eqnlabel\@vacuum}  } 
 
\def\preprint{\twocolumn\sloppy\flushbottom\parindent 1em 
        \leftmargini 2em\leftmarginv .5em\leftmarginvi .5em 
        \oddsidemargin -.5in    \evensidemargin -.5in 
        \columnsep 15mm \footheight 0pt 
        \textwidth 250mmin      \topmargin  -.4in 
        \headheight 12pt \topskip .4in 
        \textheight 175mm 
        \footskip 0pt 
        \def\@oddhead{\thepage\hfil\addtocounter{page}{1}\thepage} 
        \let\@evenhead\@oddhead \def\@oddfoot{} \def\@evenfoot{} } 
 
\def\titlepage{\@restonecolfalse\if@twocolumn\@restonecoltrue\onecolumn 
     \else \newpage \fi \thispagestyle{empty}\c@page\z@  
        \def\thefootnote{\fnsymbol{footnote}} } 
 
\def\endtitlepage{\if@restonecol\twocolumn \else  \fi 
        \def\thefootnote{\arabic{footnote}} 
        \setcounter{footnote}{0}}  
 
\catcode`@=12 
\relax 
\def\beq{\begin{equation}} 
\def\eeq{\end{equation}} 
\def\Im{\mathop{\rm Im}} 
\def\NP#1#2#3{Nucl. Phys. \underline{#1} (19#2) #3} 
 
\def\PL#1#2#3{Phys. Lett. \underline{#1} (19#2) #3} 
\def\PR#1#2#3{Phys. Rev. \underline{#1} (19#2) #3} 
 
\def\Re{\mathop{\rm Re}}

\def\crbig{\\\noalign{\vspace {3mm}}}

\relax
 
\begin{document} 
\topmargin-2.4cm
\begin{titlepage} 
\begin{center} 
\hfill{NEIP--99--002} \\
\hfill{hep-th/9902107} \\
\hfill{February 1999} \\
\end{center} 
\vspace{1.5cm}
\begin{center}{\Large\bf 
An Effective Supergravity for the Thermal Phases \\[2mm]
of N=4 Strings } 
\vskip .4in 
{\bf J.-P. Derendinger$^{\star}$} 

\vspace{.6cm}
Institut de Physique \\ 
Universit\'e de Neuch\^atel \\ 
CH--2000 Neuch\^atel, Switzerland 
\end{center} 

\vspace{1.1cm}
\begin{center} 
{\bf Abstract} 
\end{center} 
\begin{quote} 
A universal effective supergravity Lagrangian 
describing the thermal phases of heterotic strings on $T^4\times S^1$, IIA and
IIB strings on $K^3\times S^1$ is constructed.
The resulting non-perturbative phase structure is discussed. 
\end{quote} 

\vspace{1.5cm}
\begin{center}
{\it To appear in the Proceedings of the 6th Hellenic School and Workshop
on Elementary Particle Physics, Corfu, Greece, September 1998.}
\end{center}
\vspace{.9cm}

\begin{flushleft}
\rule{8.1cm}{0.2mm}\\
$^{\star}$
{\small Research supported in part by
the European Union under the TMR contract ERBFMRX-CT96-0045
and by the Swiss Office for Education and Science.} 
\end{flushleft}

\end{titlepage} 
\setcounter{footnote}{0}
\setcounter{page}{0}
\newpage 
%
%
\section{Introduction} 

String theories develop instabilities at high temperature $T$.
The $T$-dependent mass of some winding states becomes tachyonic at the
Hagedorn temperature.
This phenomenon has been described \cite{T2, AK, T3}
in perturbative string theories by identifying 
the (modular invariant) modification at finite temperature of the string 
partition function. The mechanism is similar to a $S^1$ compactification of
(euclidean) time with radius $R=(2\pi T)^{-1}$, supplemented
by a spin-dependent periodicity phase related, because of modular 
invariance, to a 
modification of the GSO projections. Supersymmetry spontaneously
breaks by a stringy Scherk-Schwarz mechanism.

In five dimensions, heterotic strings on $T^4\times S^1$, IIA and IIB strings
on $K_3\times S^1$ are related by $S$- and $T$-dualities. 
An extension of the perturbative description of strings at finite 
temperature with a universal (duality-invariant) temperature 
modulus should then display an interesting non-perturbative 
structure of thermal phases.
This contribution reviews some aspects of this analysis, as performed in
ref. \cite{ADK} (where a more complete list of references can be found). 
Other aspects are described in the contribution of Costas Kounnas
\cite{KCorfou} to these proceedings. 

The starting point of the construction is the BPS
mass formula for $N_4=4$ four-dimensional
strings, and its temperature deformation. In terms of the 
(heterotic) string coupling $g_H$ and the $T^2$ torus radii $R$ and $R_6$, 
the supersymmetric mass formula is \cite{N=4BPS}:
\beq
\label{mass4}
\begin{array}{rcl}
{\cal M}^2 &=& \displaystyle{
\left[ {m\over R}+{nR\over\alpha_H^\prime}
+ g_H^{-2}\left({\tilde m^\prime\over R_6}+{\tilde n^\prime
R_6\over\alpha_H^\prime} \right) \right]^2
+\left[{m^\prime\over R_6}+{n^\prime R_6\over\alpha_H^\prime}
+ g_H^{-2}\left({\tilde m\over R}+{\tilde nR\over\alpha_H^\prime}
\right)\right]^2}
\crbig
&=& \displaystyle{{\left|
m+ntu +i (m^\prime u + n^\prime t) + is \left[
\tilde m + \tilde n tu - i(\tilde m^\prime u +\tilde n^\prime
t)\right]
\right|^2 \over \alpha_H^\prime tu}\,.}
\end{array}
\eeq
The integers $m,n,m^\prime, n^\prime$ are the
four electric momentum and winding numbers for the four
$U(1)$ charges from $T^2$ compactification. The numbers $\tilde m,
\tilde n, \tilde m^\prime, \tilde n^\prime$ are their magnetic
non-perturbative partners, from the heterotic point of view.
In the finite temperature case, 
the radius $R$ is identified with the inverse temperature, 
$R=(2\pi T)^{-1}$. The mass formula reads then:
\beq
\label{mass5}
{\cal M}^2_T =
\left({m+Q'+{kp\over 2}\over R}+
k~T_{p,q,r}~R\right)^2-2 ~T_{p,q,r}~\delta_{|k|,1}
{}~\delta_{Q',0}\, ,
\eeq
where we have set $m'=n'={\tilde m}={\tilde n}=0$ to retain only 
the lightest states and $Q^\prime$ is the (space-time) helicity charge.
The integer $k$ is the common divisor of 
$(n,{\tilde m^\prime},{\tilde n^\prime})\equiv k(p,q,r)$
and $T_{p,q,r}$ is an effective string tension
$$
T_{p,q,r}={p\over\alpha_H^\prime}
+{q\over\lambda_H^2\alpha_H^\prime}
+{r R_6^2\over\lambda_H^2(\alpha_H^\prime)^2}\, ,
$$
with $\lambda_H^2 = g_H^2 RR_6/\alpha_H^\prime$ (six-dimensional
heterotic string coupling). 
Note that $\tilde m^\prime=kq$ corresponds to the wrapping number of
the heterotic five-brane around $T^4\times S^1_R$, 
while $\tilde n^\prime=kr$ corresponds to the same
wrapping number after performing a T-duality along the $S^1_{R_6}$
direction, which is orthogonal to the five-brane. All winding 
numbers $n, {\tilde m^\prime},
{\tilde n^\prime}$ correspond to magnetic charges from the field
theory point of view. Their masses are proportional to the
temperature radius $R$ and are not thermally shifted.

A nicer expression of the effective string tension $T_{p,q,r}$ is:
\beq
\label{Tpqr}
T_{p,q,r}={p\over\alpha_H^\prime}
+{q\over\alpha_{IIA}^\prime}
+{r\over\alpha_{IIB}^\prime}\,,
\eeq
where $\alpha_H^\prime = 2\kappa^2 s$, 
$\alpha_{IIA}^\prime = 2\kappa^2 t$ and  
$\alpha_{IIB}^\prime = 2\kappa^2 u$ 
when expressed in Planck units. 

The mass formula (\ref{mass5}) possesses the same duality properties
as the zero-temperature expression (\ref{mass4}). In addition, $R$,
the inverse temperature, is a duality-invariant quantity.
Eq. (\ref{mass5}) gives the states and 
critical values of the temperature
radius at which a tachyon appears. Each corresponds to the 
Hagedorn transition of a perturbative string, either heterotic, 
or IIA or IIB. It also contains new information (on critical 
values of $\lambda_H$ and/or $R_6$) since it also
decides which tachyon arises first when $T\sim 1/R$ increases. 

\section{Four-dimensional effective supergravity}\label{seceff}

Five-dimensional strings at finite temperature
can effectively be described by a four-dimensional supergravity,
in which supersymmetry is spontaneously broken by thermal effects.
Since we want to limit ourselves to the description of instabilities,
it is sufficient to only retain, in the full $N_4=4$ spectrum, the
potentially massless and tachyonic states, as given by mass
formula (\ref{mass5}). We can then consider only spin 0 and 1/2 states
in chiral multiplets coupled to $N_4=1$ supergravity\footnote{The 
four gravitinos remain degenerate at
finite temperature.}. 

The scalar manifold of a generic, {\it unbroken}, $N_4=4$ theory is
\cite{DF}--\cite{FK}
\beq
\label{manif1}
\left({Sl(2,R) \over U(1)}\right)_S \times \, G/H, \qquad\qquad
G/H =
\left({SO(6,r+n)\over SO(6)\times SO(r+n)}\right)_{T_I,\phi_A}.
\eeq
The manifold $G/H$ of the vector multiplets
splits into a part that includes the $6r$
moduli $T_I$, and a second part with the infinite number
$n\rightarrow\infty$ of BPS states $\phi_A$.

In the manifold $G/H$, we are only interested in keeping the six BPS
states $Z_A^\pm$, $A=1,2,3$, which
generate thermal instabilities in heterotic,
IIA and IIB strings. For consistency\footnote{See formula (\ref{mass5})
which depends on {\it three} moduli $s=\Re S$, $t=\Re T$ and $u
=\Re U$.}, these states must be
supplemented by two moduli $T$ and $U$ among the $T_I$'s. We consider
heterotic and type II strings respectively on $T^4\times S^1_6\times
S^1_5$ and $K_3\times S^1_6\times S^1_5$, where $S^1_6$ is a trivial
circle and $S_5^1$ is the temperature circle. The moduli $T$ and $U$
describe the $T^2\equiv S^1_5\times S^2_6$ torus. Thus, $r+n = 8$ in
the $N_4=4$ manifold (\ref{manif1}). To construct the appropriate
truncation of the scalar manifold $G/H$,  
we use a $Z_2\times Z_2$ subgroup contained in the $SO(6)$ 
R-symmetry of the coset $G/H$ for projecting out
non-invariant states of the $N_4=4$ theory with $r+n=8$.

A single $Z_2$ would split $H=SO(6)\times SO(8)$ in $[SO(2)\times
SO(2)]\times[SO(4)\times SO(6)]$, and the scalar manifold would
become
\beq
\label{manif2}
\begin{array}{l}
\displaystyle{\left({Sl(2,R) \over U(1)}\right)_S \times
\left({Sl(2,R) \over U(1)}\right)_T \times
\left({Sl(2,R) \over U(1)}\right)_U \times
\left({SO(4,6)\over SO(4)\times SO(6)}\right)_{\phi_A}.}
\end{array}
\eeq
At this stage, the theory would have $N_4=2$ supersymmetry and the
first three factors in the scalar manifold are vector multiplet
couplings with prepotential ${\cal F}=iSTU/X_0$. The last one is a
quaternionic coupling of hypermultiplets. The second $Z_2$ projection
acts on this factor and reduces it to
\beq
\label{manif3}
\left({SO(2,3)\over SO(2)\times SO(3)}\right)_{Z_A^+}\times
\left({SO(2,3)\over SO(2)\times SO(3)}\right)_{Z_A^-},
\qquad A=1,2,3.
\eeq
This is a K\"ahler manifold for chiral multiplets coupled to $N_4=1$
supergravity. The second $Z_2$ projection also truncates
$N_4=2$ vector multiplets into $N_4=1$ chiral multiplets.

The structure of the truncated scalar manifold and the Poincar\'e $N_4=4$
constraints on the scalar fields \cite{dR} indicate that the
K\"ahler potential can be written as \cite{AK, N=4}
\beq
\label{Kis}
K = -\log(S+S^*)-\log(T+T^*)-\log(U+U^*) 
-\log Y(Z_A^+,Z_A^{+*}) -\log Y(Z_A^-,Z_A^{-*}),
\eeq
with
\beq
\label{Yis}
Y(Z_A^\pm,Z_A^{\pm*}) = 1 -2Z_A^\pm Z_A^{\pm*} + (Z_A^\pm
Z_A^\pm)(Z_B^{\pm*}Z_B^{\pm*}).
\eeq
This K\"ahler function can be determined for instance 
by comparing the gravitino mass terms in the
$N_4=1$ Lagrangian and inthe  $Z_2\times
Z_2$ truncation of $N_4=4$ supergravity.

The $N_4=1$ effective supergravity is defined by $K$ and a holomorphic 
superpotential $W$. And $N_4=4$ supergravity is defined 
once a field representation
of the scalar manifold has been found, by specifying a 
{\it gauging} \cite{dR} which in particular
generates the scalar potential and gravitino mass terms. Finding the correct
gauging for finite temperature $N_4=4$ strings involves some guesswork
\cite{AK, N=4} and its translation into the superpotential is again easily
obtained by comparing gravitino mass terms. For mass formula (\ref{mass5}),
we find:
\beq
\label{Wis}
W = 2\sqrt2 \left[ {1\over2}(1-Z_A^+Z_A^+)(1-Z_B^-Z_B^-)
+\,(TU-1)Z_1^+Z_1^- + SUZ_2^+Z_2^- +
STZ_3^+Z_3^-  \right].
\eeq
Notice that the same effective Lagrangian could have been 
derived at the $N_4=2$ level,
using a single $Z_2$ truncation of the theory. 

Eqs. (\ref{Kis}--\ref{Wis}) define a supergravity theory which
includes in its solutions the thermal phases of five-dimensional $N_4=4$
heterotic, type IIA and type IIB strings. It possesses all duality 
properties expected from these theories. We now turn to
a survey of the most interesting phases. 

\section{Masses in the low-temperature phase}

The $N=1$ supergravity scalar potential generated by the K\"ahler
function (\ref{Kis}) and the superpotential (\ref{Wis}) can be
written
\beq
\label{Apot1}
V =
\kappa^{-4}e^K\,\Delta, \qquad\qquad
\Delta = \Delta_S+\Delta_T+\Delta_U+\Delta_++\Delta_-.
\eeq
Each contribution $\Delta_S,\Delta_T,\ldots$, to the potential 
is a polynomial in the fields $S$, $T$, $U$ and $Z_A^\pm$. In addition, $V$
only depends on {\it quadratic} combinations of $Z_A^\pm$
and their conjugates. It is then invariant under
$Z_A^\pm\rightarrow-Z_A^\pm$ and stationary at $Z_A^\pm=0$ with
respect to these fields. Since $V(S,T,U,Z_A^\pm=0)\equiv0$,
it is also stationary for all values of $S$, $T$ and $U$. 
This vacuum will be stable as long as the masses (squared) of the
fields $Z_A^\pm$ remain positive. This will be the case for 
{\it low temperature}, {\it weak string couplings} and 
{\it intermediate values} of the compactification radius $R_6$.

The low-temperature phase is common to the three strings, it
is to some extent a self-dual phase. This is also visible in the pattern of
supersymmetry breaking. Since
\beq
\label{lowTsusy}
({\cal G}^S_S)^{-1/2}\,{\cal G}_S =
({\cal G}^T_T)^{-1/2}\,{\cal G}_T =
({\cal G}^U_U)^{-1/2}\,{\cal G}_U = -1
\eeq
(other derivatives of ${\cal G} = K+\log|W|^2$ vanish), 
the canonically normalized Goldstino is the sum of the fermionic partners
of $S$, $T$ and $U$. And the gravitino mass is
\beq
\label{mgavapp}
m_{3/2}^2 = \kappa^{-2}e^{\cal G} = {1\over4\kappa^2stu}
= {1\over4 R^2}
= {1\over2\alpha^\prime_H tu} = {1\over2\alpha^\prime_{IIA} su}
= {1\over2\alpha^\prime_{IIB} st}.
\eeq

The scalar and fermion mass matrices split into four
sectors: $Z_1^\pm$ (heterotic winding modes), $Z_2^\pm$ (IIA
windings), $Z_3^\pm$ (IIB windings) and $S,T,U$ (moduli). As already
mentioned, the moduli sector is trivially massless since the
potential at $Z_A^\pm=0$ is flat.

In the $Z_1^\pm$ sector, 
comparing with the perturbative heterotic mass formula, the eigenstates 
and their masses can be identified with states with momentum
and winding numbers $m$ and $n$:

\medskip
\begin{tabular}{llll}
${1\over2}\Re(Z_1^++Z_1^-):\quad$ & mass$^2=
{1\over2\alpha^\prime_H}\left[tu+(tu)^{-1}-6\right],\quad$ &
$m=\pm1,\quad$ & $n=\mp1$; \crbig
${1\over2}\Im(Z_1^+-Z_1^-):$ & mass$^2=
{1\over2\alpha^\prime_H}\left[tu+(tu)^{-1}-2\right],$ &
$m=0,$ & $n=\pm1$; \crbig
${1\over2}\Re(Z_1^+-Z_1^-):$ & mass$^2=
{1\over2\alpha^\prime_H}\left[tu+(tu)^{-1}+2\right],$ &
$m=0,$ & $n=\pm1$; \crbig
${1\over2}\Im(Z_1^++Z_1^-):$ & mass$^2=
{1\over2\alpha^\prime_H}\left[tu+9(tu)^{-1}-2\right],$ &
$m=\pm1,$ & $n=\pm1$.
\end{tabular}
\medskip

\noindent
The first state only can become tachyonic if $(\sqrt2-1)^2 
\le tu \le (\sqrt2+1)^2$.
The existence of an upper and a lower bound with inverse critical values of
$tu$ is a manifestation of heterotic temperature duality (a $T$-duality). 

In the $Z_2^\pm$ sector, the eigenstates are perturbative IIA states with
momentum and winding numbers $m$ and $\tilde m^\prime$:

\medskip
\begin{tabular}{llll}
${1\over2}\Re(Z_2^++Z_2^-):\quad$ & mass$^2=
{1\over2\alpha^\prime_{IIA}}\left[su-4\right],\quad$ &
$m=0,\,\,$ & $\tilde m^\prime=\pm1$; \crbig
${1\over2}\Im(Z_2^++Z_2^-):$ & mass$^2=
{1\over2\alpha^\prime_{IIA}}\left[su+4(su)^{-1}\right],$ &
$m=\pm1,$ & $\tilde m^\prime=\pm1$; \crbig
${1\over2}\Im(Z_2^+-Z_2^-):$ & mass$^2=
{1\over2\alpha^\prime_{IIA}}\left[su+4(su)^{-1}\right],$ &
$m=\pm1,$ & $\tilde m^\prime=\pm1$; \crbig
${1\over2}\Re(Z_2^+-Z_2^-):$ & mass$^2=
{1\over2\alpha^\prime_{IIA}}\left[su+4\right],$ &
$m=0,$ & $\tilde m^\prime=\pm1$.
\end{tabular}
\medskip

\noindent
Again, the first state only is tachyonic if $su < 4$. 
Finally, in the $Z_3^\pm$ sector, the scalar mass matrix is 
obtained by replacing $u$ by $t$ in the
$Z_2^\pm$ mass matrix, as a result of IIA--IIB duality. The
discussion of the mass spectrum is then similar for string states
with momentum number $m$ and IIB winding $\tilde n^\prime$. Again,
thermal instabilities are generated in the field direction
${1\over2}\Re(Z_3^++Z_3^-)$ only, if $st<4$.

Using the variables 
\beq\label{xisare}
\xi_1 = tu = {2R^2\over\alpha^\prime_H}, \qquad
\xi_2 = su = {2R^2\over\alpha^\prime_{IIA}}, \qquad
\xi_3 = st = {2R^2\over\alpha^\prime_{IIB}},
\eeq
suggested by mass formula (\ref{mass5}), the vacuum with $Z_A^\pm=0$
is stable when
\beq
\label{Tis}
T = {1\over 2\pi\kappa} \left({1\over stu}\right)^{1/2}
< {(\sqrt2-1)^{1/2}\over4\pi\kappa}.
\eeq
It is then the {\it low-temperature phase}.
Since the (four-dimensional) heterotic, IIA and IIB couplings are
respectively
$$
s = \sqrt2 g_H^{-2}, \qquad t = \sqrt 2 g_A^{-2}, \qquad
u = \sqrt2 g_B^{-2},
$$
this phase exists in the perturbative regime of all three strings.
The relevant light thermal states are just the massless modes of the
five-dimensional $N_4=4$ supergravity, with thermal mass scaling like
$1/R \sim T$.

\section{High-temperature heterotic phase}\label{secmass}

This phase is defined by
\beq
\label{hhp1}
\xi_H > \xi_1 > {1\over\xi_H},
\qquad\qquad \xi_2>4, \qquad\qquad \xi_3>4,
\eeq
with $\xi_H =(\sqrt2+1)^2$. The domain of the moduli space that avoids 
type II instabilities is defined by the inequalities $\xi_{2,3}>4$.
In heterotic variables,
\beq
\label{Tineq}
2\pi T < {1\over 2\alpha^\prime_H g_H^2}\, {\rm min}\left(
R_6\,\,;\,\,
\alpha^\prime_H/R_6\right)
= {1\over 4\sqrt2\kappa^2}\, {\rm min}\left( R_6\,\,;\,\,
\alpha^\prime_H/R_6\right).
\eeq
Type II instabilities are unavoidable when $T>T_{\rm self-dual}$,
with
$$
2\pi T_{\rm self-dual} = {1\over 2g_H^2\sqrt{\alpha^\prime_H}}
= {2^{1/4}\over 4\kappa g_H}.
$$
The high-temperature heterotic phase cannot be reached\footnote{From
low temperature.} for any value of the radius $R_6$ if
the (lowest) heterotic Hagedorn temperature is higher than 
$T_{\rm self-dual}$, which translates into
\beq
\label{glim}
g_H^2 > g^2_{\rm crit.}= {\sqrt2+1\over2\sqrt2} \,\sim \,0.8536.
\eeq
Only type
II thermal instabilities exist in this strong-coupling regime and the
value of $R_6/\sqrt{\alpha^\prime_H}$ decides whether the type IIA or
IIB instability will have the lowest critical temperature.
If on the other hand the heterotic string is weakly coupled,
$ g_H<g_{\rm crit.}$,
the high-temperature heterotic phase is reached for values of the
radius $R_6$ verifying
\beq
\label{R6ineq}
2\sqrt2g_H^2(\sqrt2-1) < {R_6\over\sqrt{\alpha^\prime_H}} <
{1\over2\sqrt2g_H^2(\sqrt2-1)}.
\eeq
The large and small $R_6$ limits, with fixed coupling $g_H$, again
lead to either type IIA or type IIB instability.

In the region of the moduli space defined by Eqs. (\ref{hhp1}) 
and after minimization with respect to 
$Z_A^\pm$, the potential becomes
$$
\kappa^4 V = -{1\over s}{(\xi_1+\xi_1^{-1}-6)^2\over
16(\xi_1+\xi_1^{-1})}.
$$
It has a stable minimum for fixed $s$ (for fixed $\alpha^\prime_H$)
at the minimum of $\xi_1+\xi_1^{-1}$:
\beq
\label{highhetmin}
\xi_1=1 , \qquad
\kappa^4 V= -{1\over 2s}.
\eeq
In units of $\alpha^\prime_H$, the temperature is fixed:
$\xi_1=1=2R^2/\alpha^\prime_H = R^2/(\kappa^2 s)$.
The transition from the low-temperature vacuum is due to a
condensation of the heterotic thermal winding mode $\Re(Z_1^++Z_1^-)$, 
or equivalently by a condensation of type IIA NS five-brane in the type
IIA picture.

At the level of the potential only, this phase exhibits a runaway
behaviour in $s$. The effective supergravity is
solved by a background with a dilaton linear in a space-like 
direction\footnote{Type II instabilities, for $\xi_{2,3}<4$,
do not stabilize the moduli $s$, $t$ and $u$. Finding a 
background for these phases left aside here is a more complicated task.}. 
Its content in terms of non-critical
strings is described in detail in ref. \cite{ADK} \footnote{See 
also ref. \cite{KCorfou}.}. We will here only
consider its mass pattern and the breakdown of supersymmetry. 

The high-temperature heterotic phase only exists in the
perturbative domain of the heterotic string, where $s$ is the
dilaton, and, by duality, in non-perturbative type II regimes.
Accordingly, supersymmetry breaking arises from $s$: 
only ${\cal G}_S$ is non-zero. 
The mass spectrum naturally splits in
two sectors, the heterotic dilaton multiplet and secondly all other
chiral multiplets $(y_a,\chi_a)$ 
that play a passive role in supersymmetry
breaking. 

\vspace{3mm}
\noindent{$\bullet$\,\, \it Fermions $\chi_a$:}

\noindent
Since the K\"ahler metric is diagonal, ${\cal G}_b=0$ and
${\cal G}^S_{cb}=0$, the mass matrix simplifies to
\beq
\label{M1/21}
({\cal M}_{1/2})_{ab} = \kappa^{-1}\,e^{{\cal G}/2}\,
({\cal G}^{-1/2})^c_a {\cal G}_{cd}({\cal G}^{-1/2})^d_b =
m_{3/2}\,({\cal G}^{-1/2})^c_a {\cal G}_{cd}({\cal G}^{-1/2})^d_b.
\eeq
Mixings can only arise from non-zero values of ${\cal G}_{ab}$ due to
superpotential contributions. Since $W$ includes a term proportional
to $TUZ_1^+Z_1^-$, these four fields, which are non-zero at the
vacuum, get mixed. Masses are completely determined (in
$m_{3/2}$ units) since all parameters are fixed in this sector. On
the other hand, fermion masses in the $Z_2^\pm$ sector are
$m_{3/2}[su\pm 1]$ and $m_{3/2}[st\pm1]$ in the $Z_3^\pm$ sector.

\vspace{3mm}
\noindent{$\bullet$\,\, \it Scalars $y_a$:}

\noindent

Scalar masses arise from Lagrangian mass terms and also from a universal mass
shift due to the linear dilaton background. The physical scalar mass 
matrix reads:
\beq
\label{M01}
\begin{array}{rcl}
{\cal M}_0^2 &=&\displaystyle{
m_{3/2}^2
\left([{\cal G}^{-1/2}]^a_e \,\, [{\cal G}^{-1/2}]^f_c\right)
\left(\begin{array}{cc}
{\cal G}^{en} {{\cal G}^{-1}}^r_n {\cal G}_{rg} &
-2sW^{-1}W^{ehS} \crbig
-2sW^{-1}W_{fgS} &
{\cal G}_{fm} {{\cal G}^{-1}}^m_p {\cal G}^{ph}
\end{array}\right)
\left( \begin{array}{c}
[{\cal G}^{-1/2}]^g_b  \crbig
[{\cal G}^{-1/2}]^d_h \end{array} \right).}
\end{array}
\eeq
Comparing with the fermion mass matrix (\ref{M1/21}), one observes
that the spectrum would be supersymmetric\footnote{In the sense of
equal boson and fermion masses.} without the off-diagonal term
proportional to $2sW^{-1}W^{Sij}$. Since supersymmetry breaks in the
$S$ direction, these contributions generate
O'Raifeartaigh-type bosonic mass shifts for states that couple in the
superpotential to $S$: these are the heterotic dyonic states
$Z_2^\pm$ and $Z_3^\pm$, which generate type II
instabilities. Heterotic perturbative
states have a supersymmmetric spectrum \cite{AK}.
The supersymmetry-breaking contributions to the scalar mass
spectrum imply the existence of non-perturbative modes lighter than
their fermionic partners. Explicitly, 
mass eigenvalues in the $Z_2$ sector are
$$
m^2_{3/2}[(su-1)^2 \pm 2su], \qquad\qquad
m^2_{3/2}[(su+1)^2 \pm 2su],
$$
to be compared with the fermion masses $|su-1|m_{3/2}$ and
$|su+1|m_{3/2}$. The mass pattern in the $Z_3^\pm$ sector is obtained
by substituting $u$ for $t$ in the $Z_2^\pm$ sector.

To summarize, the spectrum is supersymmetric in the perturbative
heterotic and moduli sector ($T,U,Z_1^\pm$), and with O'Raifeartaigh
pattern in the non-perturbative sectors:
$$
\begin{array}{rrcl}
Z_2^\pm:&  m_{bosons}^2 &=& m_{fermions}^2 \pm 2su\, m_{3/2}^2,
\crbig
Z_3^\pm:&  m_{bosons}^2 &=& m_{fermions}^2 \pm 2st\, m_{3/2}^2.
\end{array}
$$

This phenomenon persists in the five-dimensional type IIA [and type
IIB] limit, in which the
$Z_3^\pm$ [$Z_2^\pm$] states become superheavy and decouple while the
$Z_2^\pm$ [$Z_3^\pm$] scalar masses are shifted by a non-perturbative
amount, since in this limit $su=1/\lambda_H^2$ [$st=1/\lambda_H^2$].
Thus, supersymmetry appears broken by non-perturbative effects.
Note, however, that this statement may not hold in the case of
the four-dimensional background solution with a dilaton motion in one
direction. This case has only an effective three-dimensional
Poincar\'e invariance, which {\it does not} imply in general mass
degeneracyin massive multiplets, even if local supersymmetry is
unbroken \cite{W, BBS}.

\vskip 1.3cm
\noindent{\bf Acknowledgements}
\vskip .1cm

\noindent
This report is based on work done in collaboration with I.
Antoniadis and C. Kounnas. I wish to thank the
organizers of the ``6th Hellenic School and Workshops on Elementary
Particle Physics'', the European Union (contracts TMR-ERBFMRX-CT96-0045 and
-0090) and the Swiss Office for Education and Science
for financial support.

\end{document}